\documentclass[lettersize,journal]{IEEEtran}
\usepackage{amsmath,amsfonts}
\usepackage{algorithmic}
\usepackage{algorithm}
\usepackage{array}
\usepackage[caption=false,font=normalsize,labelfont=sf,textfont=sf]{subfig}
\usepackage{textcomp}
\usepackage{stfloats}
\usepackage{url}
\usepackage{multirow}
\usepackage{verbatim}
\usepackage{graphicx}
\usepackage{cite}
\usepackage{booktabs}
\usepackage{bbding} 
\usepackage{graphicx}
\usepackage{makecell}
\usepackage[subfigure]{tocloft}
\usepackage{pifont}
\usepackage{threeparttable}
\usepackage{xcolor} 
\begin{document}

\title{Channel-Aware Multi-Domain Feature Extraction
for Automatic Modulation Recognition in MIMO Systems}

\author{Yunpeng Qu, Yazhou Sun, Bingyu Hui, Jintao Wang,~\IEEEmembership{Senior Member,~IEEE,}

and Jian Wang,~\IEEEmembership{Senior Member,~IEEE}

\thanks{
This work is supported by the BNRist projects (No.BNR20231880004 and No.BNR2024TD03003). \textit{(Corresponding author: Jian Wang.)}

Yunpeng Qu, Yazhou Sun, Bingyu Hui, Jintao Wang, and Jian Wang are with the Department of Electronic Engineering, Beijing National Research Center for Information Science and Technology (BNRist), Tsinghua University, Beijing 100084, China (e-mail: qyp21@mails.tsinghua.edu.cn; sunyz22@mails.tsinghua.edu.cn; huiby23@mails.tsinghua.edu.cn; wangjintao@tsinghua.edu.cn; jian-wang@tsinghua.edu.cn).
}
}



\maketitle

\begin{abstract}
Automatic modulation recognition (AMR) is a key technology in non-cooperative communication systems, aiming to identify the modulation scheme from signals without prior information. 
Deep learning (DL)-based methods have gained wide attention due to their excellent performance, but research mainly focuses on single-input single-output (SISO) systems, with limited exploration for multiple-input multiple-output (MIMO) systems.
The confounding effects of multi-antenna channels can interfere with the statistical properties of MIMO signals, making identification particularly challenging.
To overcome these limitations, we propose a Channel-Aware Multi-Domain feature extraction (CAMD) framework for AMR in MIMO systems.
Our CAMD framework reconstructs the transmitted signal through an efficient channel compensation module and achieves a more robust representation capability against channel interference by extracting and integrating multi-domain features, including intra-antenna temporal correlations and inter-antenna channel correlations.
We have verified our method on the widely-used dataset, MIMOSig-Ref, with complex mobile channel environments. 
Extensive experiments confirm the performance advantages of CAMD over previous state-of-the-art methods.
\end{abstract}


\section{Introduction}
\IEEEPARstart{A}{utomatic} modulation recognition (AMR) is one of the most critical technologies in non-cooperative communication systems and cognitive radio \cite{khan2017cognitive}. 
It finds wide applications in military and civilian domains, such as electronic warfare, resource optimization, and spectrum management.

The objective of AMR is to identify the modulation scheme of received signals in complex environments. 
Traditional methods can be categorized into two classes: likelihood-based \cite{chavali2011maximum} and feature-based approaches \cite{aslam2012automatic}.
However, their reliance on prior information and poor generalization limits applicability in non-cooperative environments \cite{qu2024enhancing}.
Recently, deep learning-based AMR methods have been widely applied. 
Works based on convolutional neural network (CNN) \cite{o2018over}, long short-term memory (LSTM) \cite{hong2017automatic}, and transformers \cite{zheng2024real} extract deep features directly from radio frequency (RF) signals or constellations, achieving results superior to traditional methods.

Most current AMR methods are designed for single-input single-output (SISO) systems. However, multiple-input multiple-output (MIMO) technology is widely deployed for its excellent spectral efficiency.
In MIMO systems, the confounding effects of the multi-antenna channel can disrupt the statistical characteristics of signals and degrade the recognition \cite{turan2015joint}.
Existing SISO-based algorithms may struggle to capture the channel states in MIMO systems. Therefore, many specialized MIMO-based AMR methods have been proposed and achieve superior results.
Huynh-The et al. \cite{huynh2022mimo} utilize three-dimensional spatial convolution to capture the correlations in multi-antenna channels.
\cite{wang2020automatic} and \cite{an2022series} utilize zero-forcing and independent component analysis (ICA) methods to combine and reconstruct multi-antenna signals, respectively.

\begin{figure}[t]
\centering
\includegraphics[width=0.83\linewidth]{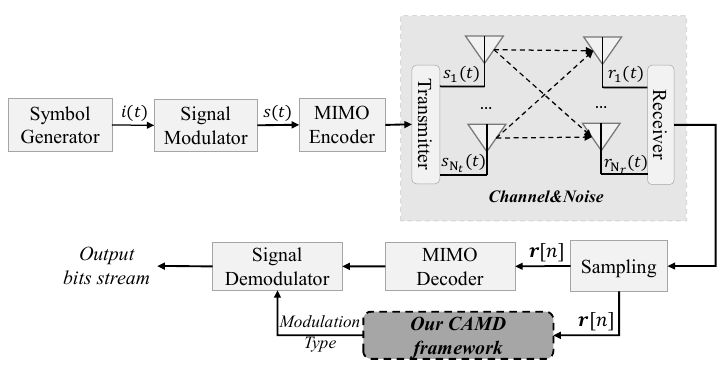}
\caption{Multiple-input multiple-output (MIMO) communication system model.}
\label{fig:system}
\end{figure}
However, these methods rely on manually defined patterns for signal reconstruction, which lack flexible perception and adaptation to complex channel states, making it difficult to capture the temporal and spatial characteristics across multiple antennas.
In complex and low signal-to-noise ratio (SNR) environments, overcoming channel interference among multiple antennas in MIMO systems and extracting discernible features from signals becomes particularly challenging.
To overcome those limitations, it is necessary to perceive and compensate for the complex channel interference and effectively integrate multi-domain features encompassing both intra-temporal and inter-antenna correlations to combat channel distortions and obtain robust representations.

In this paper, we propose a novel \textbf{C}hannel-\textbf{A}ware \textbf{M}ulti-\textbf{D}omain feature extraction (CAMD) framework for modulation recognition in MIMO systems.
We introduce an adaptive channel compensation module to preprocess the received multi-antenna signals, aiming to compensate for interference in complex environments and reconstruct the transmitted signals.
To better extract modulation characteristics of MIMO signals in the spatio-temporal domain, we propose a feature extractor based on the transformer and LSTM blocks, aiming to capture both the inter-antenna spatial correlations and the intra-antenna temporal correlations.
Our CAMD integrates channel sensing and multi-domain feature fusion within a unified framework, offering a robust solution for AMR of MIMO signals in complex scenarios.
Our contributions are as follows:
\begin{enumerate}
\item{
We propose CAMD, a novel framework for AMR in MIMO systems.
Our CAMD utilizes a well-designed feature extractor to capture both intra- and inter-antenna correlations, thereby mitigating channel interference.
}
\item{
We design an adaptive channel compensation module to further eliminate channel distortion and reconstruct the statistical properties of the transmitted signal.
}
\item{
Our CAMD significantly outperforms the baseline methods, particularly in complex environments with low SNR.
Extensive ablation experiments validate the effectiveness of various components.
}
\end{enumerate}

\section{System model}
\begin{figure*}[t]
\centering
\includegraphics[width=0.92\linewidth]{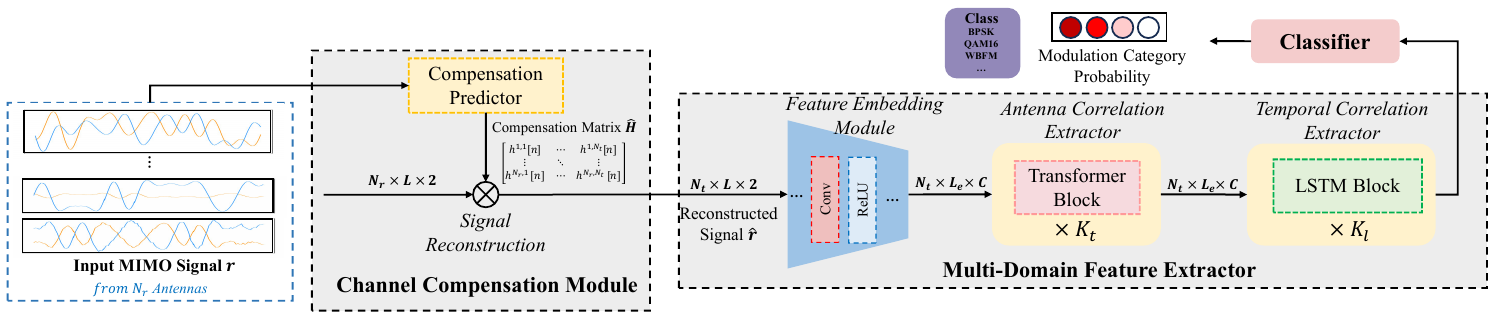}
\caption{Diagram of our proposed \textbf{C}hannel-\textbf{A}ware \textbf{M}ulti-\textbf{D}omain feature extraction (CAMD) framework. 
The input signal is processed by a channel compensation module, which predicts a compensation matrix and reconstructs the transmitted signal. Subsequently, the reconstructed signal utilizes a feature extractor to extract and integrate multi-domain features for obtaining the final prediction. }
\label{fig:model}
\end{figure*}

A typical MIMO communication system model is represented in the form of Fig. \ref{fig:system}.
At the transmitter, source-generated symbol stream $i(t)$ is modulated to $s(t)$, then sent via $N_t$ parallel transmit antennas using an orthogonal basis.
Transmitted signals $\boldsymbol{s}(t)=[s_{1}(t), s_{2}(t), ..., s_{N_t}(t)]^T$ from different antennas propagate through the wireless channel and to $N_r$ receiving antennas, where $N_r \geq N_t$.

The received signals $r(t)$ can be represented as follows:
\begin{equation}
\begin{aligned}
\boldsymbol{r}(t) & =\boldsymbol{H}(t) \boldsymbol{s}(t) + \boldsymbol{N}(t).
\end{aligned}
\end{equation}
where $\boldsymbol{r}(t)=[r_{1}(t), r_{2}(t), ..., r_{N_r}(t)]^T$.
$\boldsymbol{N}(t)$ is the additive white Gaussian noise (AWGN).
$\boldsymbol{H}(t) \sim \mathcal{C} \mathcal{N}\left(0, \boldsymbol{I}_{N_r}\right)$ is the frequency-domain channel impulse response matrix, with its elements $H^{j,i}(t)$ representing the channel coefficient between the $j$-th receiving antenna and the $i$-th transmitting antenna.

The continuous signals $\boldsymbol{r}(t)$ are sampled with length $L$ and then undergo the Hilbert transformation to the discrete matrix with both in-phase and quadrature (I/Q) components $\boldsymbol{r} = [\boldsymbol{r}^I, \boldsymbol{r}^Q] \in \mathbb R^{N_r \times L \times 2}$.
Our CAMD framework takes $\boldsymbol{r}$ as input and predicts the modulation type, thereby providing information for subsequent demodulation processes for $\boldsymbol{r}$.

\section{Methods}
\subsection{Problem Definition and Framework Overview}
MIMO-based AMR is modeled as a classification problem.
For the MIMO signal $\boldsymbol{r}$, the core design of CAMD is to learn a mapping function $f: \mathbb R^{N_r \times L \times 2} \to p(c_{gt}|\boldsymbol{r})$ that maximizes the probability of the groundtruth modulation scheme $c_{gt}$.

In Fig.~\ref{fig:model}, signal $\boldsymbol{r}$ first enters the channel compensation (CC) module to predict the compensation matrix $\hat{\boldsymbol{H}}$ and reconstruct the signal $\hat{\boldsymbol{r}}$, where the whole process is defined as mapping $\mathcal{\xi}(\cdot)$.
$\hat{\boldsymbol{r}}$ is then fed into the extractor $\phi$ to extract multi-domain deep features, and outputs the logit vectors $\boldsymbol{z} \in \mathbb R^{K}$.
\begin{equation}
\begin{aligned}
\boldsymbol{z} = \phi(\hat{\boldsymbol{r}}) = \phi(\xi({\boldsymbol{r}}))=[z_1, z_2,& ..., z_K]^T,\\
\hat{p}(c | \boldsymbol{r})  = \frac{\exp(z_c)}{\sum\nolimits_{c \in \mathcal{C}} \exp(z_c)}, \\
\end{aligned}
\end{equation}
where $K=|\mathcal{C}|$. $z_c$, $\hat{p}(c | \boldsymbol{r})$ is the logit and predicted probability for each category $c$ in the modulation set $\mathcal{C}$, respectively.

During training, we adopt the cross-entropy loss to make the predictions fit the label distribution $p(c|\boldsymbol{r})$, where $p(c|\boldsymbol{r})=1$ if $\boldsymbol{r}$ belongs to category $c$ and zero otherwise.
Our CC module and extractor are jointly optimized in an end-to-end manner.
\begin{equation}
 \begin{split}
\min \ \mathcal{L} = -\sum\nolimits_{c \in \mathcal{C}} p(c|\boldsymbol{r}) \log(\hat{p}(c|\boldsymbol{r})).\\
\end{split}
\end{equation}
During inference, we directly find the predicted modulation type $\hat{c}$ with the highest probability.
We use average recognition accuracy as the evaluation metric for different SNR scenarios.

\subsection{Channel Compensation Module}
MIMO systems face inherent channel complexity, including inter-antenna interference and distortions like fadings and offsets, which distort signals and blur modulation-related features.
Thus, we propose the CC module to correct channel-induced distortions and restore signal integrity for recognition.

Firstly, we use a compensation predictor $\zeta(\cdot)$ to predict the compensation matrix $\hat{\boldsymbol{H}}$, which should be the pseudo-inverse of the channel state matrix $\boldsymbol{H}$ in an ideal scenario, allowing us to recover the original signal $s(t)$ through complex-domain multiplication to eliminate the influence of the channel state.
\begin{equation}
\begin{aligned}
\zeta(\boldsymbol{r}) & = \hat{\boldsymbol{H}} = [\hat{\boldsymbol{H}}_I, \hat{\boldsymbol{H}}_Q] \in \mathbb R^{N_r \times N_t \times L \times 2}.\\
\end{aligned}
\end{equation}

Subsequently, we compensate the channel state based on the prediction results $\hat{\boldsymbol{H}}$ to reconstruct the transmitted signal $\hat{\boldsymbol{r}} = [\hat{\boldsymbol{r}}^I, \hat{\boldsymbol{r}}^Q] \in \mathbb R^{N_t \times L \times 2}$.
The computations are as follows:

\begin{equation}
\begin{aligned}
\hat{\boldsymbol{r}}_i^I &= \sum_{j=1}^{N_r} \hat{\boldsymbol{H}}_I^{j, i}\otimes
\boldsymbol{r}^I_j - \hat{\boldsymbol{H}}_Q^{j, i}\otimes \boldsymbol{r}^Q_j, \\
\hat{\boldsymbol{r}}_i^Q &= \sum_{i=1}^{N_r} \hat{\boldsymbol{H}}_Q^{j, i}\otimes
\boldsymbol{r}^I_j + \hat{\boldsymbol{H}}_I^{j, i}\otimes\boldsymbol{r}^Q_j.\\
\end{aligned}
\end{equation}
where $\otimes$ refers to the element-wise product. 
$\hat{\boldsymbol{r}}_i^I, \hat{\boldsymbol{r}}_i^Q$ are the reconstructed IQ signal from the $i$-th transmitting antenna.
$\hat{\boldsymbol{H}}_I^{j, i}, \hat{\boldsymbol{H}}_Q^{j, i}$ represents the compensation coefficient between the $j$-th receiving antenna and the $i$-th transmitting antenna.

The compensated signal $\hat{\boldsymbol{r}}$ is fed into the next stage to extract more robust features.
In practice, for non-cooperative scenarios with unknown transmitting antennas, we assume $N_t = N_r$ to avoid information loss during the reconstruction.
Our CC module is optimized end-to-end during training based on the final modulation classification loss, to adaptively compensate for channel distortions.

\subsection{Multi-Domain Feature Extractor}


Due to the complexity of MIMO channels, extracting more robust features requires fully exploring complementary discriminative information from multiple domains.
Among these, inter-antenna and temporal correlations are particularly crucial, as modulated signals exhibit inherent spatial distribution patterns across antennas, while modulation information is reflected in the dynamic changes in the time domain.

Therefore, we propose a multi-domain feature extractor to capture deep features, as illustrated in Fig.~\ref{fig:model}.
The extractor can be divided into three parts, which we will introduce separately.

\subsubsection{Feature Embedding Module}
The model begins with a CNN-based feature embedding module that extracts local features from the signal and maps them into token representations. After the first projection layer maps the IQ signal $\hat{\boldsymbol{r}}$ into
representations $\boldsymbol{X}_e^0 \in \mathbb R^{N_t \times L \times C}$, $K_c$ alternating convolutional and activation layers progressively extract feature mappings:
\begin{equation}
\begin{aligned}
\boldsymbol{X}_e^{k+1} = \delta(\mathbf{Conv}(X_e^{k})), k=0,1,...,K_c-1,
\end{aligned}
\end{equation}
where $\delta$ is the activation function.
Following \cite{qu2024enhancing}, we set the convolutional stride to $s=2$ to shorten the sequence length and improve the efficiency in subsequent modules, as local features of adjacent tokens tend to be redundant.
The final tokens are $\boldsymbol{X}_e^{K_c}  \in \mathbb R^{N_t \times L_e \times C}$ where $L_e = \lfloor \frac{L}{2^{K_{c}}}\rfloor$. 

\subsubsection{Inter-Antenna Correlation Extractor}
Integrating inter-antenna information is crucial for identifying unique spatial patterns of modulation schemes across antennas, yet previous methods mostly relied on simple weighted fusion via convolution.
In this paper, we innovatively propose an inter-antenna correlation extractor composed of $K_t$ transformer blocks, motivated by the parallel spatial relationships across different antennas.
LSTMs rely on sequential modeling, while CNNs are limited to weighted fusion within local fields, making them difficult to capture global correlations.
In contrast, the transformer's self-attention mechanism enables dynamic weight allocation across all antennas, allowing the model to amplify critical spatial correlations while suppressing noise.

We use tokens of different antennas in the same time slot as units and achieve interaction via self-attention.
\begin{equation}
\begin{aligned}
\boldsymbol{Q}=\boldsymbol{X}_t^{k}\boldsymbol{W}^Q&,\; \boldsymbol{K}=\boldsymbol{X}_t^{k}\boldsymbol{W}^K,\;
\boldsymbol{V}=\boldsymbol{X}_t^{k}\boldsymbol{W}^V, \\
Attn(\boldsymbol{X}_t^{k}) &= Softmax(\frac{\mathbf{Q}\mathbf{K}^T}{\sqrt{C}})\mathbf{V},
\end{aligned}
\end{equation}
where $\boldsymbol{X}_t^{k} \in \mathbb R^{N_t \times L_e \times C}$ denotes the features of $k$-th layer and $\boldsymbol{W}^Q, \boldsymbol{W}^K, \boldsymbol{W}^V \in \mathbb R^{C \times C}$ are projection weights. 
The attention map reflects the spatial correlations among all antennas for each time slot in the sequence.
We employ the multi-head self-attention (MHSA) to map features into different subspaces for better representational capability.
Following MHSA, a ReGLU feed-forward network (FFN) composed of linear and activation layers is used to re-map the features, expressed as:
\begin{equation}
\label{ReGLU}
FFN(\boldsymbol{X}_t^{k}) =  (\delta(\boldsymbol{X}_t^{k}\boldsymbol{W}_{f_1})\otimes (\boldsymbol{X}_t^{k}\boldsymbol{W}_{f2}))\boldsymbol{W}_{f3}.
\end{equation}
Each block has an MHSA followed by an FFN, enabling the extractor to effectively capture antenna-domain correlations.

\subsubsection{Temporal Correlation Extractor}
Key information of modulated signals, including phase shifts or frequency variations, is inherently embedded in temporal dynamics.
Thus, after the inter-antenna correlation extractor, we propose an LSTM-based temporal correlation extractor to integrate features from temporal perspectives.
As RF signals are time-series data, the sequential nature of temporal correlations aligns perfectly with the sequential modeling paradigm of LSTM.

We treat each token of the time-series sequence as a basic unit. 
LSTM leverages gating mechanisms to facilitate the storage and flow of information between different tokens, thereby extracting temporal correlation features from the signals.
The feature map from the transformer is input to the $K_l$ LSTM blocks as $\boldsymbol{X}_l^{0}\in \mathbb R^{N_t \times L_e \times C}$.
We pool the final vector of the LSTM output $\boldsymbol{X}_{l}^{K_l}$ from different antennas as the input to the classifier $\mathcal{C}(\cdot)$, resulting in the probability $\hat{p}(c|\boldsymbol{r})$:

\begin{equation}
\begin{aligned}
\boldsymbol{X}_l^{k+1} &= \mathbf{LSTM}(X_l^{k}), k=0,1,...,K_t-1, \\
&\hat{p}(c|\boldsymbol{r}) = \mathcal{C}(Pool(\boldsymbol{X}_l^{K_l}[:,-1])).
\end{aligned}
\end{equation}

In practice, we find that using features from only one antenna (\textit{e.g.}, $\boldsymbol{X}_l^{0}[m], m=0,1,...,N_t-1$) as the input to the LSTM blocks yields similar results compared to using the whole features $\boldsymbol{X}_l^{0}$.
This demonstrates the effectiveness of the inter-antenna correlation extractor, as in the integrated feature map, segments of each antenna have incorporated information from the global scope of all antennas.

\subsubsection{Design Alignment of our CC Module}
For the CC module, we argue that the extracted features are consistent with those required by the backbone extractor. 
This consistency stems from the fact that the CC module also needs to extract modulation-related features to adaptively compensate for channel interference.
Therefore, we adopt the same architecture as the multi-domain extractor with a lightweight parameterization to achieve more effective channel compensation.

\section{Experimental Results}
\subsection{Datasets and Experimental Setups}
\begin{table*}[t]
\scriptsize
\centering
\caption{Performance Comparisons of all methods}
\label{tab:mimosig}
\begin{threeparttable}
\begin{tabular}{ c | cc | ccc | cc |ccc}
\toprule
\multirow{3}{*}{\textbf{Benchmarks}} &  \multicolumn{5}{c|}{\textbf{MIMOSigRef-SD ($2 \times 2$)}} &  \multicolumn{5}{c}{\textbf{MIMOSigRef-SD ($4 \times 4$)}}\\
\cmidrule(lr){2-11}

~ &  \multirow{2}{*}{Params/FLOPs} &  \multirow{2}{*}{Lantency} & \multicolumn{3}{c|}{Accuracy(\%)} & \multirow{2}{*}{Params/FLOPs} &  \multirow{2}{*}{Lantency} & \multicolumn{3}{c}{Accuracy(\%)} \\
~ &  ~ & ~ & Max & Low\tnote{1} & Avg. &  ~ & ~ & Max & Low & Avg.\\
\midrule

ResNet \cite{o2018over} &92.89K/12.96M &3.13ms &  89.0 & 48.0 & 62.45 &92.89K/25.91M & 5.78ms &86.4 & 54.7 & 66.68 \\

FEA-T\cite{9915584}& 270.37K/8.90M &2.76ms & 79.5 & 45.2 & 56.25&270.37K/17.80M& 4.02ms&  79.7 & 49.4 & 61.08  \\

MCLDNN\cite{xu2020spatiotemporal} & 370.33K/169.84M& 26.97ms & 89.1 & 48.0 & 62.98 &370.33K/339.67M& 40.89ms & 91.4 & 56.3 & 68.98 \\
 
LSTM2\cite{hong2017automatic} &200.73K/102.24M &18.45ms& 79.8 &48.3& 60.12&200.73K/204.49M &32.40ms & 79.9 & 55.1 & 64.91 \\

TLDNN \cite{qu2024enhancing}  & 254.55K/17.93M & 3.95ms & 85.4 &47.7& 61.54 &254.55K/35.87M& 7.50ms & 85.8 & 55.2 & 66.82\\

MobileRaT \cite{zheng2023mobilerat}  & 259.93K/53.52M & 10.21ms &88.6&48.4&62.67&259.93K/107.04M& 19.70ms &86.3 & 55.5& 66.86 \\

4D2DConvNet \cite{ren2024mimo}  &305.98K/36.17M & 8.02ms &85.7&52.5&64.53&471.61K/72.34M  & 18.01ms &  88.4 & 59.7& 70.10 \\

CO-AMC \cite{wang2020deep}\tnote{2}  & 92.89K/12.96M &3.13ms & 88.3 & 48.0 & 62.51&92.89K/25.91M & 5.78ms & 89.8 & 55.0& 68.16 \\

MONet\cite{huynh2022mimo} &147.03K/111.54M & 19.97ms &75.5&49.5&59.50&147.03K/223.02M& 34.21ms & 89.7 & 56.7 & 67.07 \\

\midrule
\textbf{CAMD} &211.13K/22.20M& 5.47ms & \textbf{90.5} &  \textbf{53.7} &  \textbf{66.80} & 211.26K/40.44M & 8.10ms & \textbf{92.5} &  \textbf{66.6} &  \textbf{74.72} \\
\bottomrule
\end{tabular}
\begin{tablenotes}    
        \scriptsize               
        \item[1] We utilize the accuracy at -4dB as low SNR conditions.
        \item[2] We utilize the ResNet architecture and employ the same cooperative rules of information from all antennas as in \cite{wang2020deep}.
      \end{tablenotes} 
\end{threeparttable}
\end{table*}
\begin{table}[t]
\centering
\caption{Dataset configuration.}
\scriptsize
\label{tab:dataset}
\begin{tabular}{c | c }
    \toprule
\textbf{Parameters} & \textbf{Configuration}\\
    \midrule
& 32APSK, 16APSK, 64APSK, 16APSK-DVB-S2, \\
&32APSK-DVB-S2, 16APSK-DVB-S2X,\\
&32APSK-DVB-S2X, 64APSK-DVB-S2X,\\
&128APSK-DVB-S2X, 256APSK-DVB-S2, \\
& 16QAM-mil188, 32QAM-mil188, \\
Modulation type& 64QAM-mil188, 256QAM-mil188, \\
&16PAM, 32PAM, 64PAM, 128PAM,\\
& 256PAM, 16PSK, 32PSK, 64PSK, \\
& 128PSK, 256PSK, 16QAM, 32QAM, \\
& 64QAM, 128QAM, 256QAM, 16APSK-DVB-SH \\
    \midrule
SNR & -20:2:30 (dB)\\
Samples & 3968873(2$\times$2) / 3941558(4$\times$4)\\
\bottomrule
\end{tabular}
\end{table}

To assess practical reliability, we conduct evaluations in complex RF environments, including fadings, multipath propagation, Doppler shift, and other distortions.
We evaluate using the open-source MIMOSigRef-SD \cite{9405666} dataset, representing a vehicular mobile MIMO channel aligned with real-world in-motion environments and extreme selective attenuation.
We select the Vehicle A/B channel models, with MIMO antenna configurations set to $N_t=N_r=2$ and $N_t=N_r=4$.
The first 256 sample points of each signal are selected as the input.
Dataset configurations are summarized in Tab. \ref{tab:dataset}, with a relatively uniform distribution across categories and SNRs.

We divide the datasets for training, validation, and testing by 6:2:2. 
In our CAMD, the number of layers in the feature extractor is $K_c=K_t=L_l=2$ with the feature channel $C=64$ for all modules.
The CC module uses the same number of layers but with $C=32$ configured, using temporal-averaged $\hat{\boldsymbol{H}}$ for reconstruction in each time slot.
Convolutional kernel size is 3, and all activation layers utilize the ReLU.
The channel dimension of the FFN is set to $2C$.
We use AdamW with an initial learning rate of 2e-3 and weight decay of 1e-3.
The model is trained with a batch size of 512 for 50 epochs.

We compare our CAMD with several state-of-the-art (SOTA) AMR baselines, including ResNet \cite{o2018over}, LST M2 \cite{hong2017automatic}, MCLDNN \cite{xu2020spatiotemporal}, FEA-T \cite{9915584}, MobileRaT \cite{zheng2023mobilerat} and TLDNN \cite{qu2024enhancing}.
Besides, we compare three other MIMO-based AMR methods: CO-AMC \cite{wang2020deep}, 4D2DConvNet \cite{ren2024mimo}, and MONet \cite{huynh2022mimo}.
All baselines are trained and tested on an NVIDIA 3090 GPU under identical settings for a fair comparison.

 \subsection{Comparison of Recognition Accuracy with Baselines}
\begin{figure}[t]
\centering
\includegraphics[width=\linewidth]{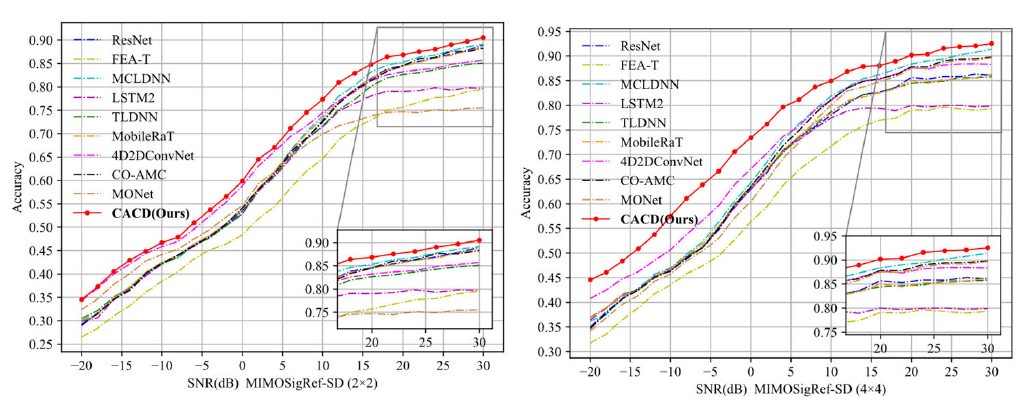}
\caption{Performance comparison with other methods.}   
\label{fig:mimosig}
\end{figure}
We compare our CLASP with other methods, with results in Tab.~\ref{tab:mimosig} and Fig.~\ref{fig:mimosig}.
Our CAMD outperforms SOTA methods across all SNR levels and achieves the highest average accuracy of 66.80\% and 74.72\% in the $2\times2$ and $4\times4$ scenarios of MIMOSigRef-SD dataset, far surpassing other methods.


Particularly, CAMD exhibits significant superiority in low SNR scenarios, where noise and channel interference have submerged the modulation-related features.
CAMD consistently outperforms other methods significantly under various low SNR conditions in Fig.~\ref{fig:mimosig}.
We attribute CAMD’s superior performance to its two core designs.
First, its CC module uses channel state information to mitigate complex channel interference and suppress noise.
Second, its multi-domain feature extractor prioritizes noise-robust and modulation-related features for retained discriminative power under severe noise.
Their complementary noise-mitigation mechanism enables CAMD to outperform rivals in low-SNR scenarios.

Furthermore, CAMD  achieves far greater improvement in the $4\times4$ scenario than the $2\times2$ scenario, which further validates our motivation.
As more antennas lead to a more complex channel environment, benefiting from the awareness of channel states and the utilization of multi-domain features, CAMD gains greater advantages in such complex settings.

\subsection{Complexity Analysis}
We utilize the widely-used metrics, including the number of parameters and floating-point operations (FLOPs), to compare the complexity.
We also provide the latency on an NVIDIA 3090 GPU for inference speed evaluation.

In Tab.~\ref{tab:mimosig}, although CAMD has achieved significant performance improvements, this does not imply a heavy additional computational burden.
Thanks to the feature mapping module shortening adjacent redundant sequences, the computational complexity of CAMD is at a similar magnitude to that of other lightweight networks, significantly lower than methods such as MCLDNN, LSTM2, and MONet.
This demonstrates its mobile-deployable potential in practical applications.


\subsection{Effectiveness of Channel Compensation}

\begin{table}[t]
\caption{Ablation on the Channel Compensation Module}
\scriptsize
\label{tab:ce}
\centering
\begin{threeparttable}
\resizebox{\linewidth}{!}{
\begin{tabular}{ c| c |c | ccc }
\toprule
\multirow{2}{*}{\textbf{CC Module}}  & \multirow{2}{*}{\textbf{Params/Flops\tnote{1}}} & \multirow{2}{*}{\textbf{Domain}} &  \multicolumn{3}{c}{\textbf{MIMOSigRef-SD ($4 \times 4$)}}\\
~ &~&~  & \textbf{Maximum} & \textbf{LowSNR} & \textbf{Average} \\
\midrule
w/o CC  & 168.80K/29.41M & - & 92.2 &  64.0 &  73.15\\
JADE & 168.80K/29.41M & - & 92.2 &  64.3 &  73.29\\
\midrule
CNN & 173.06K/30.30M & Local  & \textbf{92.9} &  64.9 &  73.80 \\
Transformer& 194.88K/36.11M  & Inter-Antenna  & 92.3 &  65.6 &  74.38\\
LSTM &  189.44K/34.63M & Temporal  & 92.2 &  65.5 &  74.04\\
\textbf{CAMD(Ours)} & 211.26K/40.44M & Multi-Domain & 92.5 &  \textbf{66.6} &  \textbf{74.72}   \\
\bottomrule
\end{tabular}}
\begin{tablenotes}    
        \scriptsize               
        \item[1] We report the parameter count and complexity of the whole neural network.
      \end{tablenotes} 
\end{threeparttable}
\end{table}

To validate the CC module, we compare with other scenarios: (1) without the CC module; (2) using the traditional JADE \cite{an2022series} for blind equalization.
In Tab.~\ref{tab:ce}, compared to other scenarios, using the CC module can improve performance by 2-3\%, showcasing the adaptive perception of channel state through deep learning modules and better compensation for channel interference compared to traditional methods.

We believe that compensating for channel state also requires the utilization of inter-antenna and temporal correlations. Hence, the CC module shares the architecture of the multi-domain extractor.
As shown in Tab.~\ref{tab:ce}, multi-domain feature extraction has achieved better results compared to using only LSTM or transformer for single-domain correlation extraction, or basic architectures like CNN to extract local features.
Furthermore, we also find that a lightweight CC module is sufficient for channel compensation without introducing excessive computational complexity.


\subsection{Effectiveness of Multi-Domain Features}


\begin{table}[t]
\caption{Ablation on the Multi-Domain Features}
\scriptsize
\label{tab:feature}
\centering
\resizebox{\linewidth}{!}{
\begin{tabular}{ c | cc | ccc }
\toprule
\multirow{2}{*}{\textbf{Extractor}}  & \multicolumn{2}{c|}{\textbf{Feature Domain}} &  \multicolumn{3}{c}{\textbf{MIMOSigRef-SD ($4 \times 4$)}}\\
~& \textbf{Inter-Antenna} & \textbf{Temporal}  & \textbf{Maximum} & \textbf{LowSNR} & \textbf{Average} \\
\midrule
CNN & \XSolidBrush & \XSolidBrush  & 90.1 &  48.7 & 64.02\\
Transformer& \Checkmark & \XSolidBrush  & 91.4 &  65.2 &  73.28\\
LSTM & \XSolidBrush & \Checkmark  & 86.5 &  65.0 & 71.40\\
\textbf{CAMD(Ours)} &\Checkmark& \Checkmark & \textbf{92.5} &  \textbf{66.6} &  \textbf{74.72}   \\
\bottomrule
\end{tabular}}
\end{table}
To validate the integration of multi-domain features, we systematically remove different modules one by one.
We compare the scenarios of only using a single module to extract either inter-antenna correlations or temporal correlations, where we change the number of blocks to ensure consistent complexity.
Additionally, we include a baseline scenario that only uses a basic CNN for local feature extraction.

In Tab.~\ref{tab:feature}, the basic baseline for extracting local features performs the worst, while incorporating inter-antenna or temporal correlations leads to significant improvement. This indicates that modulation-related patterns manifest in global correlations.
Integrating multi-domain features results in better performance compared to single-domain features, indicating that fusing features across multiple domains is critical for obtaining more discriminative representations. 
Especially in MIMO systems, extracting the correlation between antennas helps improve the generalization ability in complex scenarios.

\section{Conclusion}
In this paper, we analyzed the challenges of modulation recognition in MIMO systems and proposed the CAMD framework.
Our CAMD utilized a multi-domain feature extractor to extract correlations from intra- and inter-antennas to combat channel interference.
In addition, we designed an adaptive channel compensation module to reconstruct the transmitted signal.
Experiments showed that CAMD achieves outstanding accuracy under complex mobile channel environments, surpassing previous SOTA methods.
\bibliographystyle{IEEEtran}
\bibliography{ref.bib}

\vfill

\end{document}